# High-frame rate, fast neutron imaging of two-phase flow in a thin rectangular channel

R. Zboray[1], I. Mor[2], V. Dangendorf[3], M. Stark[1], K. Tittelmeier[3], M. Cortesi[1,4], R. Adams[4]

[1] Paul Scherrer Institute, PSI Villigen, CH-5232, Switzerland

[2] Soreq NRC, Yavne 81800, Israel

[3] Physikalisch-Technische Bundesanstalt (PTB), 38116 Braunschweig, Germany

[4] Swiss Federal Institute of Technology Zürich, Sonnegstrasse 3, CH-8092 Zürich, Switzerland

**ABSTRACT**

We have demonstrated the feasibility of performing high-frame-rate, fast neutron radiography of air-water two-phase flows in a thin channel with rectangular cross section. The experiments have been carried out at the accelerator facility of the Physikalisch-Technische Bundesanstalt. A polychromatic, high-intensity fast neutron beam with average energy of 6 MeV was produced by 11.5 MeV deuterons hitting a thick Be target. Image sequences down to 10 millisecond exposure times were obtained using a fast-neutron imaging detector developed in the context of fast-neutron resonance imaging. Different two-phase flow regimes such as bubbly slug and churn flows have been examined. Two phase flow parameters like the volumetric gas fraction, bubble size and bubble velocities have been measured. The first results are promising, improvements for future experiments are also proposed.

1. **INTRODUCTION**

Application of neutron imaging as a non-intrusive, high resolution technique is increasingly popular in many engineering and applied science disciplines. In the field of two-phase flow thermal and cold neutron imaging has also found increasing application in the last three decades, mainly because they provide a better contrast for aqueous two-phase flows in a metallic casing in comparison to other techniques such as X- or gamma-rays. This predestinates neutron imaging for general two-phase flow research (Mishima & Hibiki, 1998). Especially in the case of high-pressure, high-temperature two-phase flows in thick metal casings. These conditions are usually found in nuclear fuel bundles, which is why most of the applications can be found in fuel bundle studies. Such studies aim to determine two-phase flow parameters with high spatial (and temporal) resolution in fuel bundle geometries, which could then be used for fuel bundle optimization either directly, based on the experimental results or indirectly by using the data to validate computational codes.

Using neutron radiography/tomography for fuel bundle research, Takenaka et al. have performed a number of studies (Takenaka et al., 1998; Takenaka & Asano, 2005) and compared the performance of fast and thermal neutron imaging for fuel bundles (Takenaka et al., 1999). Lim et al. (2005) performed high-frame rate thermal neutron radiography on a model fuel bundle. Mishima et al., (1999) have examined a model of two adjacent subchannels of a high-conversion, light water reactor tight-lattice fuel bundle using neutrons. Along this line, the research by Kureta (2007a, 2007b) and Kureta et al., (2008) must be also mentioned who have performed comprehensive investigations on tight-lattice fuel bundles using thermal neutron tomography. Recently, Zboray et al. (2011) have performed a comprehensive study of annular flows, providing an accurate measurement of the time-averaged liquid film thickness distribution with high spatial resolution using cold-neutron imaging. The technique has been thoroughly optimized (Zboray and Prasser, 2013a) and applied to different fuel bundle model geometries such as two neighboring subchannels (Kickhofel et al., 2011; Zboray et al., 2011) or four and five neighboring subchannels of a fuel bundle (Zboray and Prasser, 2013b).

Furthermore, one can profit from the penetrating power of neutrons when imaging liquid-metal two-phase flows and steam explosions (Mishima & Hibiki, 1998; Saito et al., 1999; Saito et al., 2005). Looking, however, at larger (full) bundle models featuring flow regimes with a substantial amount of water and also higher amounts of metal, cold and thermal neutron imaging becomes unfeasible due to the limited penetrating power.

In such cases, the higher penetrating power of fast neutrons could enable higher-contrast imaging.

A fast neutron imaging system is proposed and is under development by Zboray et al (2014) for fuel bundle studies and beyond. Takenaka et al. (1999) have also shown the advantages of fast neutrons with respect to thermal ones for imaging low-gas-fraction conditions in a rod bundle geometry using a beam line of a research reactor. Besides the aforementioned works, the application of fast neutrons is only mentioned in some early research papers focusing mainly on neutron scatterometer measurements (Banerjee et al., 1978; Hussein, 1988) providing time and cross-section averaged gas fractions in two-phase flow channels. Neutron scatterometer measurements are also performed recently on a 7-pin CANDU rod bundle model reported by Buell et al. (2005) using an isotopic source (Cf-252). The application of fast neutron imaging for two-phase flow research is much scarcer than that using thermal or cold neutrons. This is mainly due to much less available intense fast neutron beam facilities and the difficulties involved in fast neutron detection for imaging purposes.

In the present study, we demonstrate the feasibility of high-frame-rate, fast neutron radiography for two-phase flows using the imaging infrastructure at the fast neutron beam line of the accelerator facilities of the Physikalisch-Technische Bundesanstalt (PTB), Braunschweig, Germany. For demonstration purposes, we have examined adiabatic, air-water two-phase flows at atmospheric pressures in a rectangular, relatively thin channel at different flow regimes, including bubbly, slug and churn flows. More specifically we aimed to measure the volumetric gas fraction, the bubble sizes and bubble rise velocities. The next section will be focused on a detailed description of the experimental setup consisting of the fast neutron beam, the imaging detector system and the two-phase flow channel. Finally we illustrate the results through a few examples and discuss the possibilities for future improvements.

## 2. THE EXPERIMENTAL SETUP

*2.1. The fast neutron beam line at PTB*

The experiments have been carried out at the high intensity, polychromatic fast neutron beam line at PTB (the spectrum is shown in Fig. 4a). The neutron beam is produced by the d+Be reaction, hitting a 5 mm thick water-cooled and wobbling Beryllium target by a pulsed deuteron beam, accelerated by a compact cyclotron to 11.5 MeV. The useful neutron energy ranges from ca 2 MeV up to 15 MeV; the fluence-averaged neutron beam energy is around 5.5 MeV. The deuteron spot size on the Be target which is equivalent to the size of the neutron source, is around $D=5$ mm in diameter. The Be target is enclosed by a collimator which restricts the neutron beam to neutrons scattered into forward directions and defines a neutron field of appropriate size. The deuteron beam current used during the experiments was 40 µA, resulting in a total flux of around 1.3E7 n/cm$^2$s at the position of the two-phase flow channel. The detailed description of the beam line and the neutron yields and energies is given in Brede et al., (1989).

*2.2. The fast neutron imaging detectors*

The images were recorded by a third generation, multiple-frame Time-Resolved Integrative Optical Neutron (TRION) detector developed at PTB in the context of high resolution, energy-selective fast neutron resonance radiography (Dangendorf et al., 2008). It is based on a plastic-fiber scintillator screen and a two stage intensified CCD camera system. The components and layout of the detector are shown in Fig. 1. The fiber scintillator screen (BCF-12, produced by Crytur, fiber size 0.7 mm) has an active surface area of $200 \times 200 mm^2$ and is $t$=50 mm thick (in the beam direction), enabling a detection efficiency of 25.7 % at En = 6 MeV. Behind the bending mirror, a 120 mm lens is focusing the light on a position sensitive optical preamplifier (OPA), which intensifies the image from the scintillator screen as well as preserves the few-nanosecond fast timing property of the scintillator, which was important in the original neutron Time-of-Flight (TOF) applications. The latter is achieved by applying a phosphor screen on this intensifier with a fast (ns) fluorescence time. The intensified light signal from the OPA falls on a kaleidoscopic image splitter which splits the image formed on the phosphor of the OPA to a field of 9 (3 x 3) sub-images. This image splitter is coupled through a lens to a 9-fold segmented image intensifier (II) of which 8 segments can be gated independently. Each of the segments on the II photocathode views the scintillator screen and can acquire an image for an independently selectable time window with exposure times ranging down to 5 ns. The purpose of the 8-fold image splitting is to enable quasi-simultaneous recording of multiple images at different time slices, either for TOF-sensitive imaging applications or for observation of fast dynamic processes, like the flow patterns in the present application. A CCD camera records all segments simultaneously on a large area CCD chip with 16 Mpixels. Further details of the TRION detector can be found in Dangendorf et al., (2008) and Mor et al., (2011).

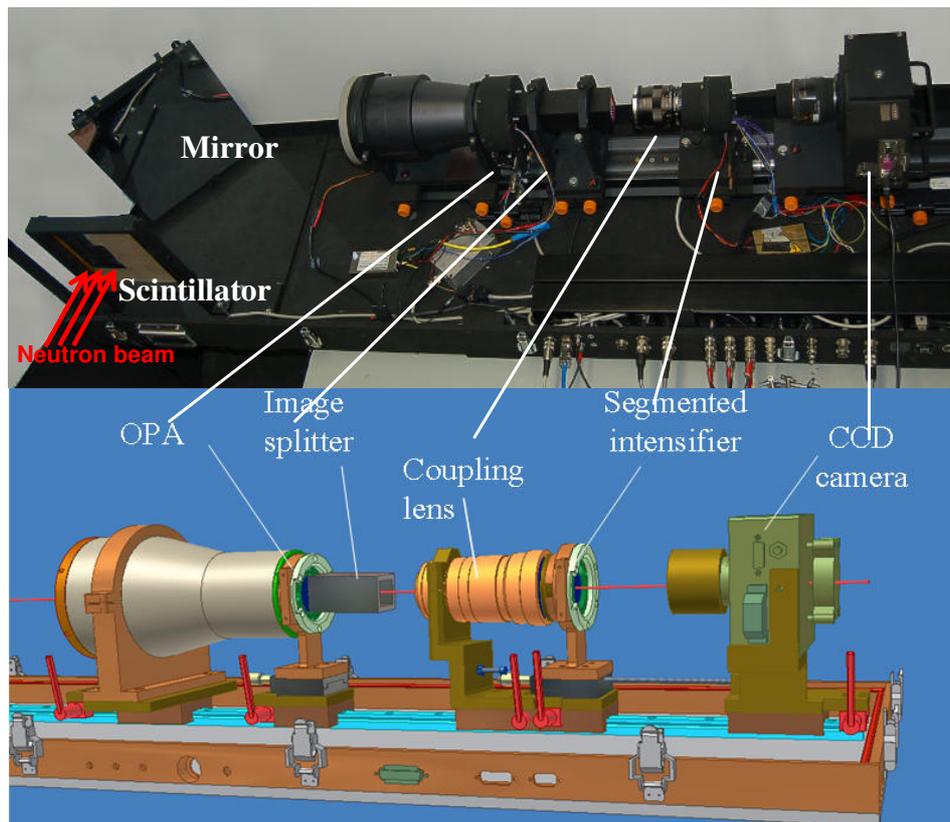

Fig. 1: The generation III TRION detector developed at PTB and used for the present study.

*2.3. The two-phase flow loop*

A thin, rectangular-shaped two-phase flow channel was constructed to investigate two-phase flows in this fast-neutron radiography. The layout of the channel is shown in Fig. 2. The channel is made of aluminum (EN AW-5083 alloy) using two half shells which are bolted together. Each shell has a wall thickness of 3mm and the total depth of the flow channel in beam direction is 15mm. This depth has been chosen to maximize the contrast for gas bubbles while still keep the channel narrow enough to avoid the overlapping of bubbles in the radiographic images. The water inlet is fed by a pump from a small water reservoir. The air inlets, featuring 5 orifices of three rows of 2, 4 and 5 mm in diameter, respectively, are fed from a pressurized air network. For these experiments, the middle row with 4mm injection orifices has been used. In the supply lines the gas and liquid flow rates, respectively are measured by rotameters with an accuracy of about 2%. The two-phase mixture exits the channel at its top, its water content flows back to the water reservoir while its air content is left to egress freely to the atmosphere. Both the water inlet and the two-phase exit have a diameter of 20 mm. The width of the flow channel is 90 mm; the total height of the channel is 1500mm. The field of view (FOV) of the imaging system is around 115x115mm$^2$. The bottom of the FOV starts around 1090 mm above the water and 1022 mm above the air inlet. This corresponds to at least 40x$D_h$ ($D_h$=4A/S is the hydraulic diameter of the channel, A being the channel cross section and S the channel perimeter) distance from the source implying a developed two-phase flow at the height of the FOV.

The channel has been placed perpendicular to the neutron beam at a distance of $L$=600 mm in front of it. The scintillator screen of the detector was at $l$=74 mm behind the center of the channel.

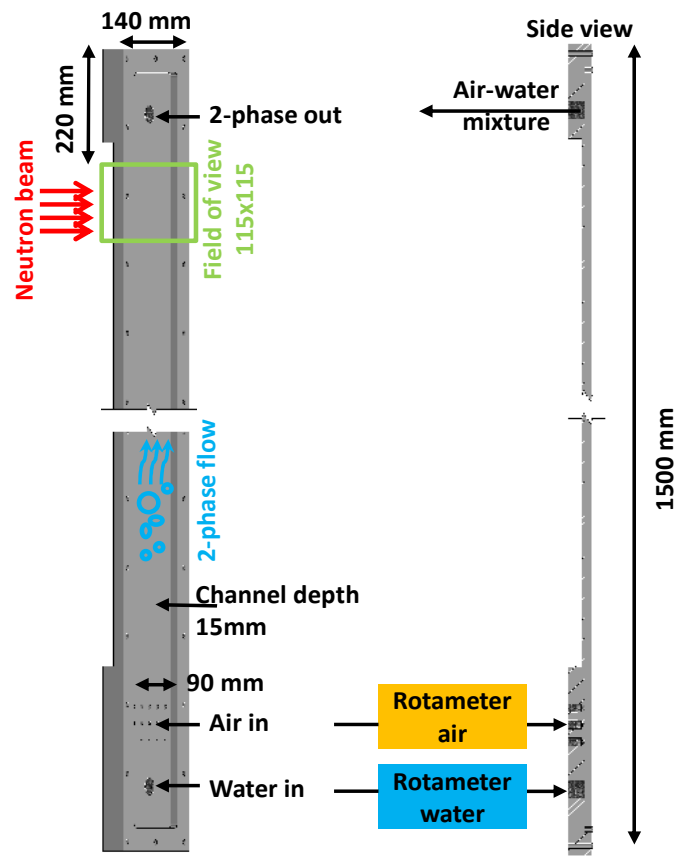

Fig. 2: Layout of one half shell of the thin, rectangular two-phase flow channel.

## 3. EXPERIMENTAL CONDITIONS

We have investigated two-phase flow regimes from bubbly to churn flow by increasing the gas flow rate. All tests were carried out at an ambient temperature of approximately 20 ºC and at atmospheric pressure (apart from some minimal pressure loss due to the flow through the channel). The two-phase flow can be characterized by the so-called superficial gas and liquid velocities:

$$J_g = \frac{\dot{V}_g}{A} \text{ and } J_l = \frac{\dot{V}_l}{A}, \tag{1}$$

where $\dot{V}$ is the volumetric flow rate. Furthermore, the gas volume flow fraction is defined as:

$$\varepsilon = \frac{\dot{V}_g}{\dot{V}_g + \dot{V}_l}. \tag{2}$$

Depending on the combination of the gas and liquid flows, the two-phase flow established in the channel shows different patterns from dispersed bubbly flow, through slug flow characterized by large Taylor bubbles, to churn and annular flow, in the order of increasing gas volume fraction. Cheng et al., (2010) give a comprehensive and recent overview on the different flow regimes in two-phase flows and their characteristics.

Table 1: Matrix of the experimental conditions.

|      | Flow regime  | Air flow [l/h] | Water flow [l/h] | $J_{air}$ [m/s] | $J_{water}$ [m/s] | ε     |
|------|--------------|----------------|------------------|-----------------|-------------------|-------|
| Exp1 | Bubbly       | 60             | 250              | 0.01            | 0.05              | 0.175 |
| Exp2 | Bubbly/Slug  | 120            | 250              | 0.02            | 0.05              | 0.297 |
| Exp3 | Slug         | 720            | 250              | 0.13            | 0.05              | 0.717 |
| Exp4 | Churn        | 2280           | 250              | 0.41            | 0.05              | 0.889 |

Table 1 summarizes the four different experimental conditions and the parameters of the corresponding flow regimes. The latter were estimated based on phenomenological flow regime maps for narrow, rectangular channels with upward two-phase flow available in the literature (see e.g. Hibiki and Mishima, 2001). Note that the transition between different flow regimes is quite smooth and thus the boundaries are not sharply defined as in the case of Exp2 in Table 1.

## 4. EXPERIMENTAL RESULTS

*4.1. Spatial resolution, FOV and contrast transfer function.*
First, we have used a special tungsten mask to estimate the contrast transfer function (CTF) of the imaging system, the spatial resolution and the image pixel size. The mask have groups of slits with varying periodicities combined with a three-step structure with increasing thickness in the beam direction as shown in Fig. 3a. The image has been taken by placing the mask directly on the front surface of the empty channel and the image of the empty channel (without mask and water filling) was used for flat-field correction. The CTF (Fig. 3b) has been estimated as described in Mor et al., (2011) by dividing the difference of the maximum and minimum gray values of the light and dark slits, respectively, with the sum of them. This is done for each group of different slit sizes giving their periodicity in line-pairs/mm (lp/mm). The CTF measures how good the system images the real contrast of the sample object at different spatial frequencies. The commonly used limit value of 10% CTF is around 0.5 lp/mm in our case. Mor et al. (2011) report higher values, however, for thinner scintillator screen and a simpler optical chain resulting in reduced noise. In our case, besides the inherent blurring effects in the detector screen (neutron scatter, recoil proton cross talk, optical cross talk, for further details see Mor et al., 2011), we also suffer from a geometrical blur due to the combination of beam spot size and a non-zero sample-to-detector distance. This, essentially a parallax, effect can be estimated as $b_2 = l/L*D$ (the definition of these quantities are found in *2.1* and *2.3*) having a value of 0.62 mm. A further significant broadening (blur) is due to the thickness of the scintillator screen. This effect grows from zero with increasing distance from the image center and is estimated as $b_1 = t*tg\alpha$, where $\alpha$ is the angle under which a given image point is seen from the neutron source spot with respect to the image center and $t$ is the thickness of the scintillator screen. Its maximum value at the corner of the FOV is $115/\sqrt{2}/600*50 = 6.8mm$. Note that it is not a normal broadening due to the attenuation of the flux inside the screen, it has rather a "comet-tail" shape. The sum of these two contributions to the total blur thus varies from 0.62 mm to 7.42 mm from the image center to its corner.

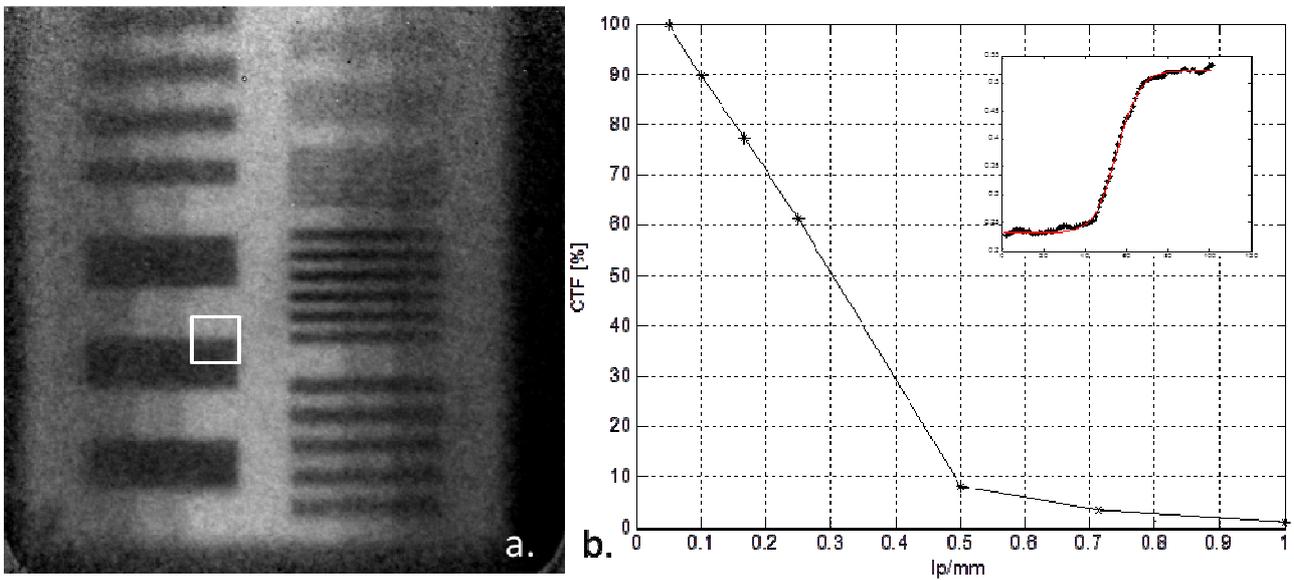

Fig. 3: Flat-field corrected image of the attenuation of the tungsten CTF mask (a) The estimated CTF of the detector system (b). The inset in (b) shows a typical average edge response function (black) and the fitted logistic function (red). The edge response is taken over the area indicated by the white rectangle in (a).

The actual spatial resolution can be estimated as the full width at half maximum (FWHM) of the derivative of the edge spread function (ESF). The ESF is obtained by taking an average ESF in the gray-scale image of the mask e.g. over an area indicated by the white rectangle in Fig. 3a. The ESF is fitted to a logistic function of the form:

$$ESF(x) = \beta_1 + \frac{\beta_2}{1+\exp[-\beta_3(\beta_4 - x)]} \qquad (3)$$

The FWHM of the derivative of such an ESF, the so-called line spread function (LSF), can be shown to be obtained as $3.53/\beta_3$. The ESF and its fitting c.f. eq. (3) are shown in the inset of Fig. 3b. FWHM=1.76 mm was obtained. The aforementioned contribution of the screen thickness and the geometrical blur is $b_1+b_2$=1.45 mm at the position of the edge where the ESF is taken (see Fig. 3a). This indicates that in our case these two components are the main contributors to the total blur and the other inherent blurring effects in the screen are less significant.

We have estimated the pixel size of our images by fitting two logistic functions to the opposite edges of the widest (1cm) slits. The difference of the centroid parameters, $\beta_4$, of the two fits was used to get a scaling factor for the images. This resulted in a pixel size of 0.1154mm. This scaling factor is also needed to account for the magnification effect caused by the divergent beam.

*4.2. Tuning of the imaging system, gain, exposure and delay time settings*
Our goal was, by setting sufficiently short exposure times, to capture individual bubbles with minimum motion blur that would allow an accurate determination of their size and volume fraction in a time-resolved manner. By sequentially repeated exposures, their motion and velocity should also be captured. As we performed two-phase flow imaging with the above introduced system for the first time, some tuning was necessary. This included among others finding optimal combinations of image exposure times, delay times between the individual II segments, and optimizing the different gains of the II and the OPA. Unfortunately, due to limited available beam time, an in-depth investigation of the effect of these parameters on the images could not be carried out. For this reason some system properties/factors that might negatively influence the results have only been revealed by post-test analysis. Two such effects are circular vignetting in the images and intensity saturation. Both are influenced by several factors but they stem mainly from the fact that the OPA was not optimized for the present application for millisecond exposures but rather for repetitive exposure cycles in the nanosecond range required for TOF imaging. For this a fast phosphor screen was used in the OPA having a low light output. This implied that the relay lens between OPA and segmented intensifier was opened to maximum aperture (small F-number), a setting which caused significant vignetting resulting in a quite non-uniform light distribution with a peak intensity for the middle image segment and with very low intensity on the peripheral segments. This effect is increased further by the use of the kaleidoscopic image splitter as the coupling of the OPA phosphor to the splitter via a relatively thick glass window suppresses the reception of the light at larger emission angles by the splitter and thus infringes the "filling" of the peripheral (far off-axis) images as compared to the central image. This effect can in principle be corrected by "flat-normalization", however neutron statistics decreases and noise deteriorates the images towards the periphery.

Another unwanted effect which cannot easily being compensated for stems from the fast, but low-light efficiency phosphor of the OPA, which was required in its original application for fast-neutron TOF imaging. The low efficiency of the phosphor was compensated for by a large gain in the MCP stages of this image intensifier. In the present application, where single shot exposure at high neutron rates was mandatory this caused gain saturation of the intensifiers. This saturation is spatially varying and depends on the impinging neutron flux and the exposure duration. Thus it cannot be easily corrected by flat normalization because the neutron flux varies locally and also between the flat (no flow) and regular exposures (with two-phase flow).

In summary the two described effects cause the central image segment to be saturated while the peripheral segments cannot utilize the full potential amplification range and are noisier. Higher gain is usually paired with lower exposure time and therefore the saturation/vignetting effects become more serious for the short exposures. Therefore such measurements are not considered here in the detailed analysis and only be used for qualitative observations and understanding of the behavior of the imaging system under the present operating conditions. For longer exposures and lower gains however, the saturation effect is much less severe and affects, due to the circular vignetting, practically only the high-intensity central segment of the 8-fold II. For those measurements, the center segment has been discarded. Note that one of the eight segments was not functioning, thus we had only six independent images to acquire in a sequence. A delay time of 5 ms between the images in the sequence has been found to work reasonably well for experiments for each flow regime. Typical exposure times where meaningful images could be obtained range from 20 to 40 ms. Several image sequences (10-15) have been taken for each experiment to obtain sufficient statistical information on the flow behavior.

*4.3. Results on two-phase flows*

To obtain the gas volume fraction in the two-phase flow, two reference (flat field) images are needed: one of the air-filled ($I_g$), and another of the water-filled channel ($I_w$). Based on that the gas volume fraction at the image position (*x,y*), averaged over the channel depth in beam (*z*) direction, can be given as:

$$\varepsilon(x,y) = \frac{-\ln\left(\frac{I(x,y)}{I_w(x,y)} \frac{D_w}{D_i}\right)}{-\ln\left(\frac{I_g(x,y)}{I_w(x,y)}\right)}. \qquad (4)$$

Where *I* is the intensity of the image taken for the two-phase flow. The dose correction factor $D_w/D_i$ is applied to compensate for small intensity fluctuations in the neutron beam between actual and reference measurements. $D_w$ and $D_i$ are mean intensity values over an area outside of the flow channel for the water-filled reference and for the actual image, respectively. In the denominator of eq. (4) no dose correction is included as the reference images have been repeated several times and then averaged, smoothing practically such dose variations out. These were usually quite small, around 2-3%, hence, the correction factor could be avoided without influencing the results much.

Before applying eq. (4) the image intensities are corrected for sample scatter and beam-hardening effects, which can bias the estimation of the gas volume fraction. Scattered neutrons from the flow channel can still reach the detector and contribute to the recorded intensity adding an undesired contribution to the detected image. The beam hardening effect is due to using a polychromatic neutron beam (see the spectrum in Fig. 4a) on a relatively strongly attenuating specimen (e.g. thick water layers). The mean energy of the transmitted neutrons is gradually shifted to higher energies, generally characterized by smaller interaction cross sections, resulting in lower apparent attenuation as compared to the expected one. Note that the energy dependent sensitivity of the fiber scintillator screen should also be taken into account in this respect. To compensate for these effects, we have performed direct Monte Carlo (MC) simulations of the imaging setup, using an approach similar to that in Zboray & Prasser (2013). The MC simulations have been performed using the general purpose MC code, MCNP Version 5 (Booth et al., 2003). The energy dependent proton conversion rate per pixel has been simulated. For a more realistic simulation of the actual detector behavior, it is then weighted by a function, $f(E_p)$, describing how many photons are created by a recoil proton of energy *E*. The latter is obtained based on the manufacturer specification of 8000 photons/1MeV electron equivalent (St. Gobain, 2011). To take into account the different light yields produced by electrons and protons in the plastic scintillator, the proton energy is converted into electron equivalent based on the work of O'Rielly et al. (1996).

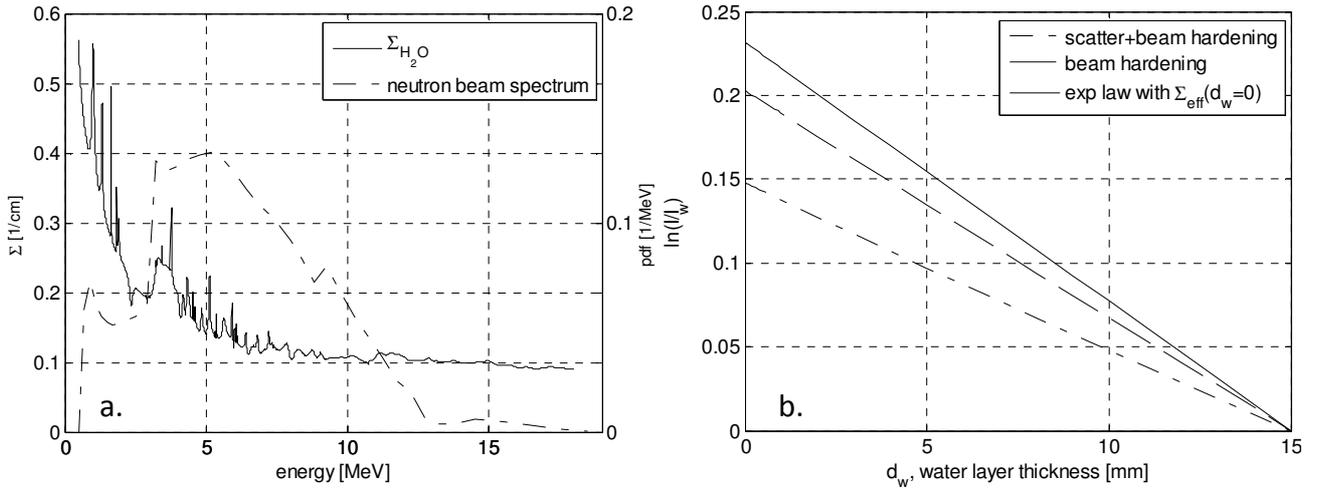

Fig. 4: (a) The polychromatic spectrum of the fast neutron beam and the macroscopic cross section of water. (b) The transmission value, $-ln(I/I_w)$, for increasing water thickness in the channel with respect to channel filled with water ($d_w$=15mm) obtained by MC simulations.

Fig. 4b shows the simulated attenuation values, $-ln(I/I_w)$, for increasing water layer thickness, $d_w$, in the channel. Clearly, in the limit $d_w=0$, both effects vanish and the larger $d_w$ is, the stronger the effects. As $d_w$ increases, the effective attenuation coefficient, $\Sigma_{H2O,eff}=-ln(I/I_w)/d_w$ that satisfies the Beer-Lambert law for the given $d_w$ is not constant but decreases due to beam hardening and the scatter contribution. A practical way to correct for these effects is:

$$-\ln\left(\frac{I}{I_w}\right)_{corr} = d_w \Sigma_{H2O,eff(d_w=0)}, \tag{5}$$

where $d_w$ is obtained from:

$$-\ln\left(\frac{I}{I_w}\right)_{MC}(d_w) = \ln\left(\frac{I}{I_w}\right)_{mes} \tag{6}$$

$-ln(I/I_w)_{MC}$ is the MC simulated attenuation as a function of $d_w$ (the dashed line in Fig. 4b). $d_w*\Sigma_{H2O,eff}(d_w=0)$ represents the transmission of the hypothetical limiting case (solid line in Fig. 4b) corresponding to the transmission law satisfied by $\Sigma_{H2O,eff}(d_w=0)$, where no beam hardening and scatter occur. The latter value is determined by extrapolating MC simulation results towards $d_w\to 0$. The correction is uniquely defined as the attenuation curves in Fig. 4b are monotonic and it is applied pixel-wise on the images. Note that the MC simulation can distinguish between un-collided and collided neutron contributions in the detector, the former being obviously free of the sample scatter, however, it still contains the beam hardening effect. This separation of the two effects shows that the scatter contribution is significantly larger (see Fig. 4b). Based on the simulation results in Fig. 4b, the effective (uncorrected) attenuation coefficient for the channel full of water is $\Sigma_{H2O,eff}(d_w=1.5cm)=0.0985$. Based on the flat field measurements with empty and water-filled channel obtained for the different detector settings we have estimated this attenuation coefficient, whose average value scatters between 0.086 and 0.105, being in relatively good agreement with the simulated one.

Some typical instantaneous volumetric gas fraction ($\varepsilon$) distributions for different two-phase flow conditions are shown in Figs 5-7. The individual bubbles for bubbly and slug flow are clearly resolved.

For the latter flow regime, note the typical elongated (~6-7cm long), large Taylor bubbles. The bubble sizes are quite large even for the bubbly flow case and correspondingly have strongly deformed (non-spherical and non-ellipsoidal) shapes. The smallest bubbles resolved are characterized by a main axes length of approximately 8-9mm as well as more ellipsoidal shape. The large bubble size is partly due to the relatively large gas-injection orifice. In future experiments, applying a more custom tailored detector variant, we might be able to resolve smaller bubbles as well.

Note that the images shown in Figs 5-7, have been denoised in two steps: first by applying a 5×5 pixel median filter. Secondly, an anisotropic diffusion based filter as proposed originally by Perona & Malik (1990) is applied. In the latter method the image intensity field is smoothed by the diffusion equation, which is solved by applying a simple forward time centralized space (FTCS) scheme. To prevent blurring of edges the diffusion is reduced there by an appropriate choice of the diffusion coefficient. The latter is made space (and thus time) dependent, more specifically an exponential function of the local image intensity gradient. This method tends to preserve even relatively small (step) edges and in certain cases can even sharpen them. The combination of these two non-linear filters eliminates much of the noise without causing a significant extra image blurring.

To compensate for the blur of the imaging system discussed in Section 4.1, one could in principle apply deblurring, edge enhancement methods. These however, in general, tend to enhance high-frequency noise significantly. For this reason we omit their application here. Given the relatively large bubble sizes observed, the blur of the imaging system might be considered less critical. In potential future experiments, with an optimized detection system allowing a better signal-to-noise ratio and resolving smaller bubbles, this issue will be revisited.

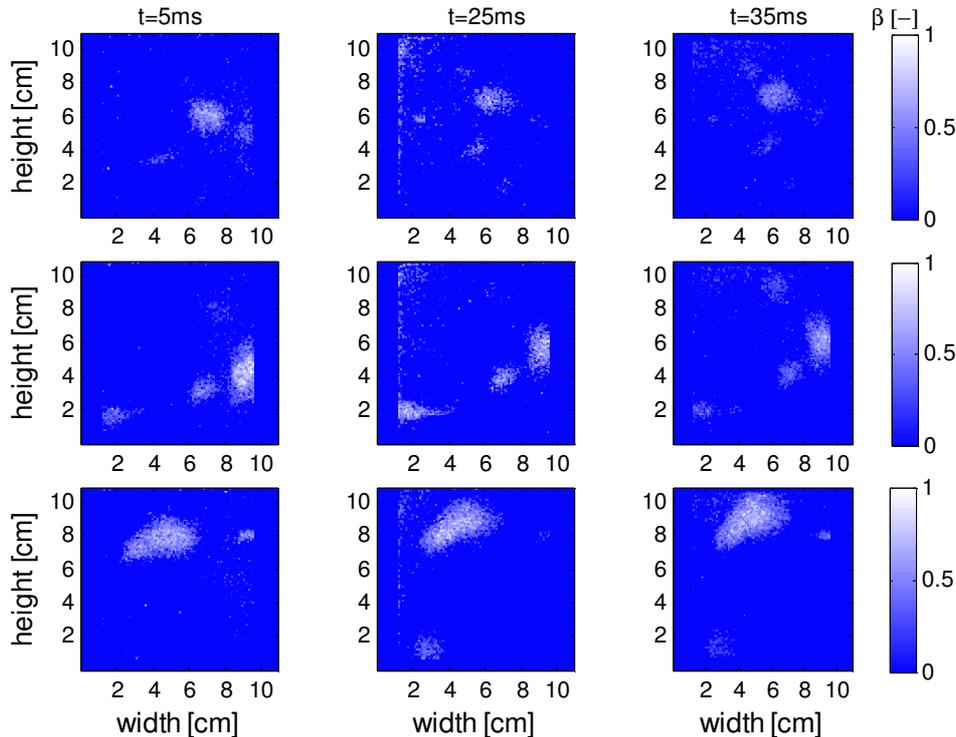

Fig. 5: Typical volumetric gas fraction distributions for bubbly flow (Exp1) conditions. Exposure time is 30ms. Only three images of the sequence of total six are shown in a row. The relative time (t=0 corresponds to the start of acquiring the first image) of the start of acquiring the images is given at the top.

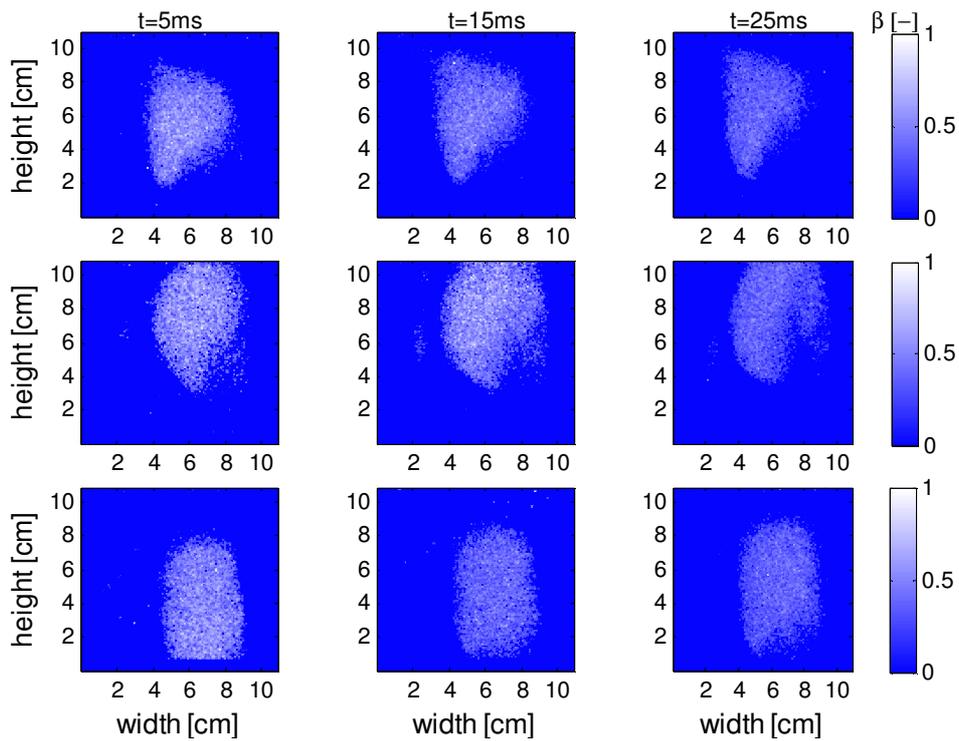

Fig. 6: Typical volumetric gas volume fraction distributions for slug flow (Exp3) conditions. Exposure time is 40ms. Only three images of the sequence of total six are shown in a row. The relative time (t=0 is the start of acquiring the first image) of the start of acquiring the images is given at the top.

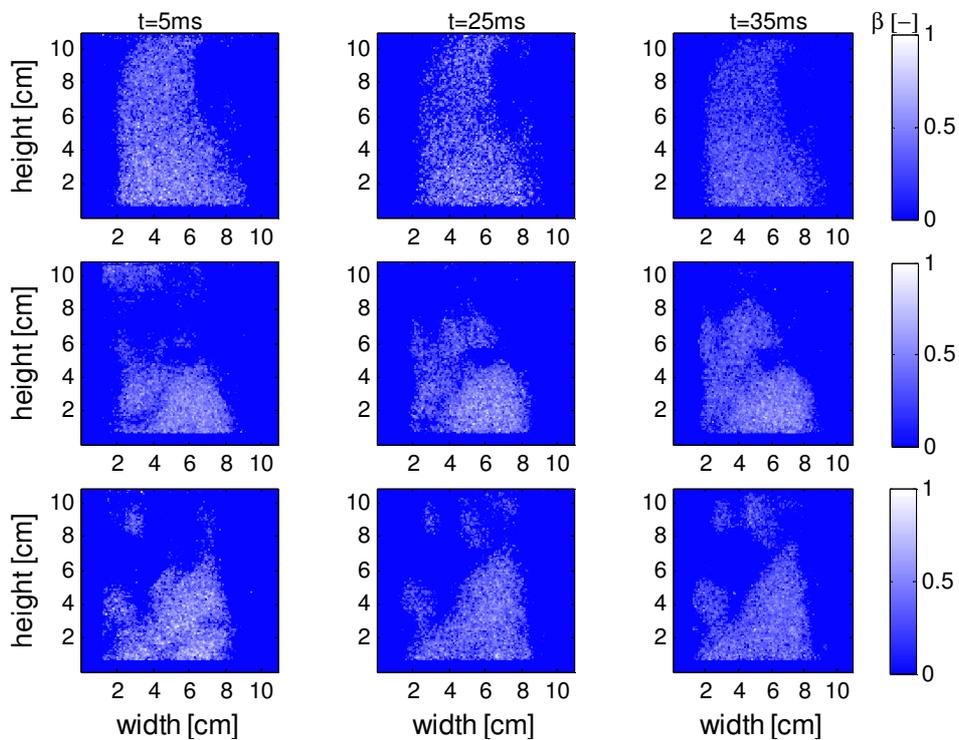

Fig. 7: Typical volumetric gas volume fraction distributions for churn flow (Exp4) conditions. Exposure time is 30ms. Only three images of the sequence of total six are shown in a row. The relative time (t=0 is the start of acquiring the first image) of the start of acquiring the images is given at the top.

Using the sequentially acquired, instantaneous gas-fraction distribution images shown in Figs. 5-7, we can try to estimate the velocity field around the moving bubbles. For this we have applied the optical flow method originally developed by the computer science community for machine-vision (Horn and Schunck, 1981). The algorithm works on the premise that for two, 2D projections of a 3D scene, there exists a two-dimensional velocity field that moves the first image toward the second. It is based on the conservation of the image brightness, $I$, at coordinates $x,y$ and time $t$:

$$\frac{DI(x,y,t)}{Dt} = 0 \tag{7}$$

Resulting in:

$$I_x u + I_y v + I_t = 0, \tag{8}$$

Where the subscripts are shorthand notation for the derivatives in that coordinate and $u,v$ are the velocity components in the $x,y$ directions, respectively. As the problem is ill-posed in this form having one equation for the two velocity components, a smoothness constraint is added as:

$$|\nabla u|^2 + |\nabla v|^2 = 0. \tag{9}$$

Usually neither the brightness nor the smoothness conditions are strictly satisfied due to imperfections in image acquisition, therefore the problem is reformulated as an optimization problem minimizing the goal function:

$$L = \int_\Omega \left[ (I_x u + I_y v + I_t)^2 + \alpha^2 \left( |\nabla u|^2 + |\nabla v|^2 \right) dxdy \right] \tag{10}$$

The weighting of the two components in the minimization function is controlled by a scalar parameter, $\alpha$. The choice of this parameter influences the result of the optical flow algorithm. We have used here an in-house implementation of the optical flow algorithm (Stark, 2013) developed to optimize the evaluation of particle image velocimetry measurements. Details of solving eq. (10) and more sophisticated strategies for choosing the $\alpha$ parameter can be found in that work. For our noisy, "speckled" images (see Fig. 5), as optical flow assumes brightness invariance between images, it also lines up the speckles between images. To get around this problem, $\alpha$ had to be chosen high enough so that it smoothes the small-scale motion of the speckles, while still accounting for the large-scale motion of the bubbles. Proposed methods to automate the selection of the smoothing parameter don't apply in this case; therefore an empirical choice has been made, where $\alpha$=0.5 was used for each image pair.

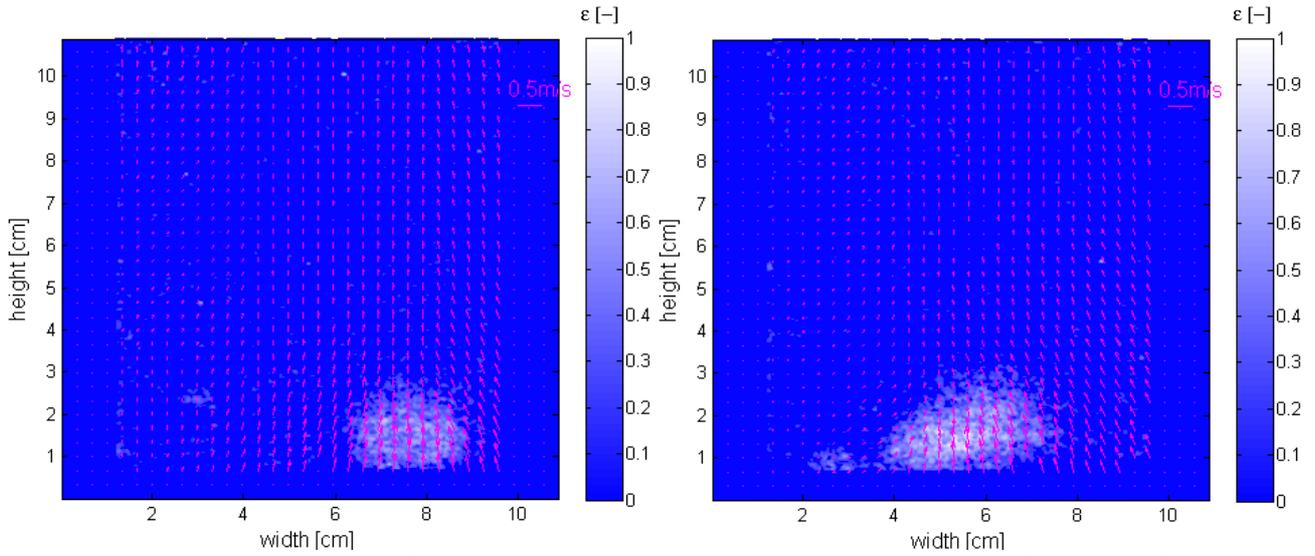

Fig. 8: Optical flow velocity field estimation around a bubble. Two examples for bubbly flow are shown. Mean bubble velocities are 0.2m/s (left) and 0.14m/s (right). Equivalent bubble diameters are 13.3mm and 19.5mm, respectively.

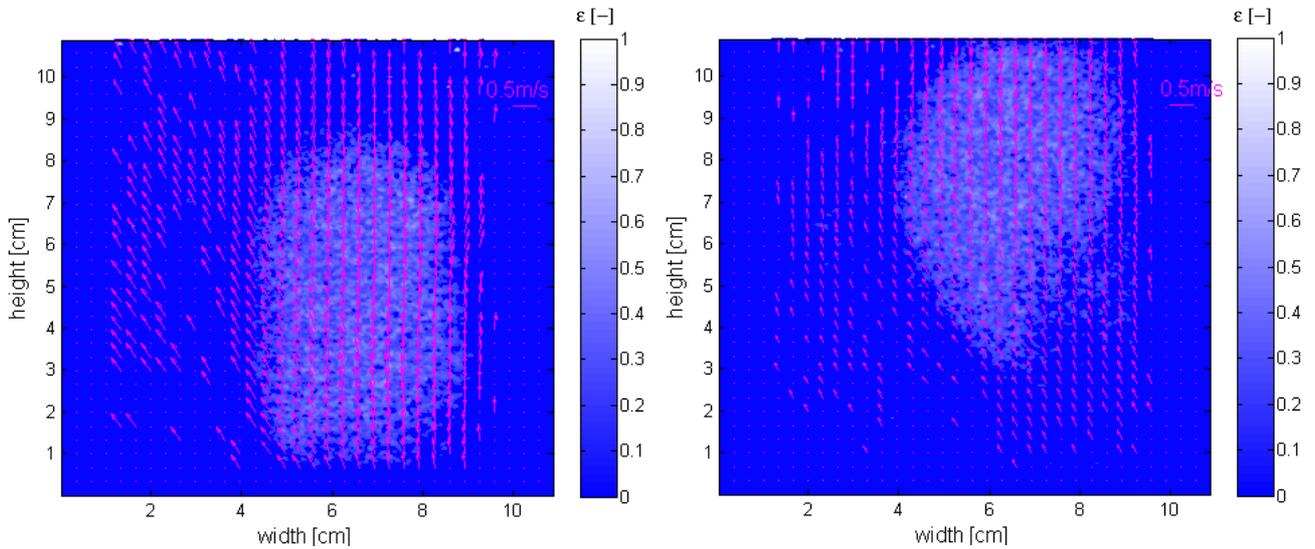

Fig. 9: Optical flow velocity field estimation around a bubble. Two examples for slug flow are shown. Mean bubble velocities are 0.37m/s (left) and 0.33m/s (right). Equivalent bubble diameters are 38.3mm and 39.2mm, respectively.

The optical flow algorithm provides a pixel-wise displacement field between two consecutive images, which is then turned to a velocity field using the time delay between the images. Figs. 8 and 9 show examples of the velocity field around bubbles determined by the optical flow method. Only the velocity vectors in and around the bubble are physically meaningful. The presence of non-zero velocity vectors further away from bubbles is due to the apparent motion of the background noise between two successive images. Additionally, optical flow, due to the smoothness term, tends to act also in regions where no intensity gradients are present, smoothly filling in information from neighboring cells.

We estimate the mean bubble velocity by averaging the velocity vectors over the area of the bubble, $A_b$. For the bubbly flow examples shown in Fig. 8, we obtain velocities around 0.15-0.2m/s. This agrees well with terminal velocity values that can be found in the literature for large, deformed air bubbles of similar equivalent diameters (see Talaia, 2007). The equivalent diameter is defined as:

$$\frac{D_e^2 \pi}{4} = A_b \qquad (11)$$

For the large slug bubbles, shown in Fig. 10, we find somewhat higher velocities around 0.35m/s. Note that at such velocities and applied exposure times the bubbles move a few millimeters during the exposure, giving rise to a motion blur. In fact, for these experiments, this is the major contributor to blur, dominating the blur of the imaging system. For this reason, as well as to be able to resolve smaller bubble sizes, it is essential to decrease exposure times down to 5-10ms at least without compromising the signal-to-noise ratio too much. This is going to be the aim of future experimental run along with optimized detector system free of vignetting and saturation effects.

## 5.   CONCLUSIONS AND OUTLOOK

We have shown the feasibility of high-frame-rate, fast neutron radiography of two-phase flow in a thin, rectangular channel using an imaging system comprised of a polychromatic fast-neutron beam and a detector based on a scintillator screen and a two-stage intensified CCD camera setup. Although the imaging detector system has been optimized for a different application, these initial results are promising. Instantaneous gas volume fraction distribution and bubble size have been measured for different flow regimes, such as: bubbly, slug and churn flows. Bubbles with diameters down to 8-9mm could be clearly observed. Using sequential recordings, the bubble velocity has been estimated using the optical flow algorithm.

Some adverse effects such as vignetting and saturation have been observed using the present detector setup. Tailoring the detector towards optimal performance for future experiments could involve the use of an optical amplifier with lower electronic (MCP) gain and a more efficient phosphor, though at an acceptable compromise regarding its speed. By replacing the present CCD camera with a high-speed CMOS camera one could also avoid using the 8-fold segmented II and the image splitter. This would eliminate both the saturation effects and vignetting.

Such a system could be used in a more optimal way for time-resolved two-phase flow radiography. It is expected that it would enable to resolve bubbles of smaller-size compared to the present study by enabling shorter exposure times, thereby minimizing motion blur.